\documentclass[12pt]{article}
\usepackage{graphicx, amsmath, amssymb, cite, setspace, color,relsize,bm,bbold,dsfont,xcolor}

\usepackage[top=1 in, bottom=1 in, left=.9 in, right=.9 in]{geometry}

\newcommand{\captionfonts}{\footnotesize} 
\makeatletter 
\long\def\@makecaption#1#2{%
  \vskip\abovecaptionskip
  \sbox\@tempboxa{{\captionfonts #1: #2}}%
  \ifdim \wd\@tempboxa >\hsize
    {\captionfonts #1: #2\par}
  \else
    \hbox to\hsize{\hfil\box\@tempboxa\hfil}%
  \fi
  \vskip\belowcaptionskip}
\makeatother

\def\fnote#1#2{\begingroup\def\thefootnote{#1}\footnote{#2}
     \addtocounter{footnote}{-1}\endgroup}

\def\gsim{ \lower .75ex \hbox{$\sim$} \llap{\raise .27ex
\hbox{$>$}} }
\def\lsim{ \lower .75ex \hbox{$\sim$} \llap{\raise .27ex
\hbox{$<$}} }

\newcommand{\blank}{\textcolor{white}{D}}

\usepackage{titlesec}

\titleformat{\paragraph}
{\normalfont\normalsize\bfseries}{\theparagraph}{1em}{}
\titlespacing*{\paragraph}
{0pt}{3.25ex plus 1ex minus .2ex}{1.5ex plus .2ex}

\begin{document}

\title{Compactifying de Sitter Naturally Selects \\ a Small Cosmological Constant}

\date{}

\author{Adam~R.~Brown,$^{1}$ Alex~Dahlen,$^{2}$ and Ali Masoumi$^{3}$ \vspace{.1 in}\\
 \vspace{-.3 em}  $^1$ \textit{\small{Physics Department, Stanford University, Stanford, CA 94305, USA}} \\
 \vspace{-.3 em}  $^2$ \textit{\small{Berkeley Center for Theoretical Physics, Berkeley, CA 94720, USA}}  \\
  \vspace{-.3 em}  $^3$ \textit{\small{Tufts Institute of Cosmology, Medford, MA 02155, USA 
}} } 

\maketitle
\fnote{}{\hspace{-.65cm}emails: \tt{adambro@stanford.edu, adahlen@berkeley.edu, ali@cosmos.phy.tufts.edu}}
\vspace{-.95cm}

\maketitle

\begin{abstract}

\noindent We study compactifications of $D$-dimensional de Sitter space with a $q$-form flux down to $D-Nq$ dimensions. We show that for $(N-1)(q-1)\geq 2$ there are double-exponentially or even infinitely many  compact de Sitter vacua, and that their effective cosmological constants accumulate at zero.  This population explosion of $\Lambda\ll1$ de Sitters arises by a mechanism analogous to natural selection.
\end{abstract}

\thispagestyle{empty} 
\newpage 
\subsubsection*{Introduction}

There's typically more than one way to compactify a higher-dimensional theory. Different ways give rise to different lower-dimensional theories with different cosmological constants. In this paper we will study the distribution of the cosmological constants of compactified  vacua.

There is a common lore that the distribution of four-dimensional de Sitter vacua has no special feature as $\Lambda \rightarrow 0^+$. And indeed this common lore 
is borne out in two well-studied models: the  Freund-Rubin (FR) model \cite{Freund:1980xh,RandjbarDaemi:1982hi,Douglas:2006es}, which has dynamical extra dimensions but only a single internal $q$-cycle wrapped by a $q$-form flux;  and the Bousso-Polchinski (BP) model \cite{Bousso:2000xa}, which has many internal $q$-cycles, each individually wrapped by a $q$-form flux,  but which fixes the geometry by fiat. In this paper we will see that the common lore does not apply when we combine these features: when we have dynamical extra dimensions compactified on a  product manifold.  

We will show that a $D$-dimensional de Sitter vacuum begets  exponentially or even infinitely many  dS$_{D-Nq}\times (S_q)^N$ vacua whose cosmological constants accumulate at zero. This population explosion arises by a mechanism analogous to natural selection. Compactifying once ($N=1$) gives a family of first generation vacua with a range of cosmological constants. Compactifying again ($N=2$) assigns a family of second generation vacua to each first generation vacuum. Since the number of offspring of a given vacuum is inversely proportional to its cosmological constant, and since the progeny all inherit a $\Lambda$ no larger than that of their parent, sequential compactification naturally selects for the trait of having a small cosmological constant. By the $N$th generation the distribution of de Sitter vacua is strongly peaked at $\Lambda =0$. 

These compactifications do not have a cosmological constant problem, in the sense that (when $(q-1)(N-1)\ge2$) the cosmological constant in a generic de Sitter vacuum is double-exponentially sub-Planckian.\footnote{\noindent Specifically, when the vacua are counted with a uniform measure over the number of flux units, we find that the generic de Sitter vacuum has minuscule cosmological constant.}  These compactifications do, however, have another problem no less severe: the KK scale in a generic vacuum  is near the Hubble scale, and is therefore also double-exponentially sub-Planckian.  The question of why the cosmological constant is so small has been replaced with the question of why the extra dimensions are so small.  Nevertheless, the accumulation point persists even when restricting to de Sitter vacua with a KK scale arbitrarily higher than the Hubble scale. 

 \subsubsection*{Bousso-Polchinski compactifications: no accumulation}
\begin{figure}[h!] 
   \centering
   \includegraphics[width=\textwidth]{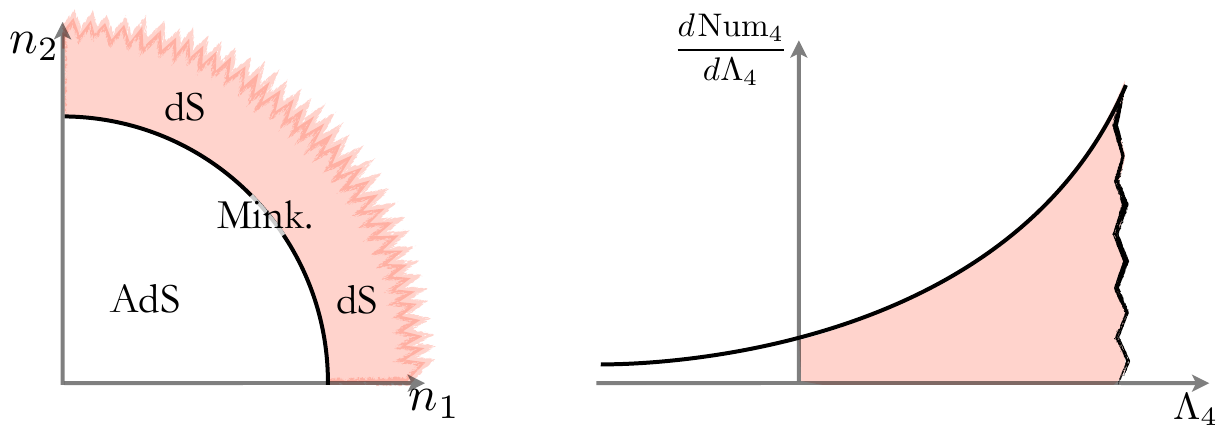} 
   \caption{The Bousso-Polchinski model. Left: the phase diagram of vacua.  The lines of iso-potential are circles/spheres. Spheres of smaller radius are AdS; spheres of larger radius are dS. If the fluxes get too large, perturbative control is lost. Right: the number density of vacua with a given value of the cosmological constant.  The total number of de Sitter minima is finite, and the number density is smooth through $\Lambda_4=0$.} 
   \label{fig:BP}
\end{figure}

 In the Bousso-Polchinski model, the extra dimensions are  fixed by fiat, and the effective four-dimensional cosmological constant is uplifted by the $q$-form flux:
\begin{equation}
\Lambda_4 = \Lambda_{\textrm{no flux}} + \frac{1}{2} \sum_{i=1}^N g^2 n_i^{\,2} ,
\end{equation}
where $g$ is the quantum of magnetic flux and $n_i \in \mathbb{Z}$ is the number of  units wrapping the $i$th $q$-cycle. 
For $\Lambda_{\textrm{no flux}} <0$  this gives rise to a landscape  with both AdS (small $n_i$) and dS (large $n_i$) vacua, but if the $n_i$'s get too large, the energy density exceeds the cutoff and perturbative control is lost. The number of de Sitter vacua that lie beneath this cutoff is exponentially large in $N$, but finite; a typical vacuum has a cosmological constant just below the cutoff.
Since nothing picks out $\Lambda=0$ as special, the  distribution of vacua  is flat through zero, as is shown in Fig.~\ref{fig:BP}.

\subsubsection*{$\bm{N=1}$ Freund-Rubin compactifications with $\bm{\Lambda_D > 0}$: no accumulation}
\begin{figure}[h!] 
   \centering
   \includegraphics[width=\textwidth]{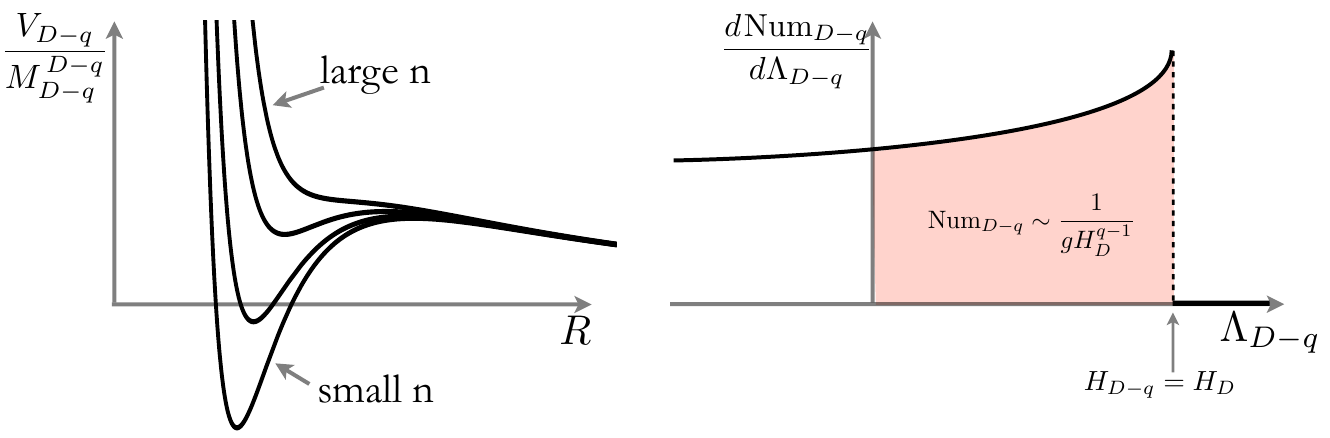} 
   \caption{The $N=1$ FR compactification of $D$-dimensional de Sitter. Left: the effective potential as a function of $R$, plotted for several values of the number of conserved flux units $n$.  Right: the number density  of dS$_{D-q}$ vacua as a function of the lower-dimensional $\Lambda_{D-q}/M_{D-q}^{\,D-q}$.  To make this histogram, we have treated $gn$ as continuous.}
   \label{fig:FR+}
\end{figure}

 Freund-Rubin compactifications  start with the Einstein-Maxwell action
\begin{equation}
S = \int d^D x \sqrt{|\det g_{\mu\nu}|} \left( M_D^{\,D-2} \mathcal{R} - \frac{1}{2q!} F_q^{\,2} - 2  \Lambda_D \right),  \label{eq:higherDaction}
\end{equation}
where $F_q$ is a $q$-form flux with $q\ge2$,  $\Lambda_D$ is a cosmological constant, and $M_D$ is the Planck mass.  We can compactify this down to $D-q$ dimensions where the internal manifold is a $q$-sphere uniformly wrapped by $n$ units of the $q$-form flux.  The $(D-q)$-dimensional Einstein-frame effective potential for the radius $R$ of the sphere is 
\begin{gather}
\frac{V_{D-q}\,(R)}{M_{D-q}^{\,D-q}} \sim \left( \frac{1}{M_D^{\,q} R_{\blank}^q} \right)^{\frac2{D-q-2}} \left( \frac{\Lambda_{D}}{M_{D}^{\,D} }  - \frac{1}{M_{D}^{\,2} R_{\blank}^2} + \frac{g^2 n^2}{M_{D}^{\,2} R_{\blank}^{2q}} \right), \label{eq:VeffN1}
\end{gather}
with the lower-dimensional Planck mass defined by $M^{\,D-q-2}_{D-q} \equiv M_{D}^{\,D-2} R_{\blank}^q$\hspace{-1mm}. The curvature term (the second term) makes the extra dimensions want to shrink, but the flux term (the third term) buttresses the extra dimensions against collapse, creating a minimum of the potential, as shown in Fig.~\ref{fig:FR+}. The value of the lower-dimensional cosmological constant is set by the value of the potential in this minimum $\Lambda_{D-q} \equiv V_{D-q}(R_\textrm{min})$. A small value of $gn$ gives rise to an AdS minimum and a larger value gives rise to a dS minimum; when $gn$ is too large, however, there is no minimum of any kind. Unlike in the BP model, the reason the minimum disappears is not that the energy density becomes too large. Instead, the minimum disappears because the flux has swelled the extra dimensions so much that they get caught up in the Hubble expansion and decompactify.  The value $n_\text{max}$ at which this occurs is that for which $R_\text{min}\sim H_D^{\,-1}$, where $H_D\equiv \Lambda_D^{1/2}/M_D^{(D-2)/2}$ is the $D$-dimensional Hubble scale.  At $n=n_\text{max}$, all three terms in the effective potential are approximately the same size, so $g^2 n_\text{max}^{\,2}\sim H_D^{-2(q-1)}.$

How many $N=1$ de Sitter minima are there?  The allowed flux values are evenly spaced in $n$, so a good proxy for the number of vacua is to treat $n$ as continuous and to evaluate the length in $n$-space.  The total number of vacua is therefore set by $\text{Num}_{D-q}\sim n_\text{max}\sim H_D^{-(q-1)}/g$.  Since $q\ge2$, the total number of lower-dimensional minima is thus inversely proportional to the higher-dimensional cosmological constant---lower de Sitter are fitter and give rise to more offspring.  The fraction of these offspring that are de Sitter is O(1) and independent of $H_D$; the distribution of their c.c.'s is smooth through $\Lambda_{D-q}=0$ and is plotted in the right pane of Fig.~\ref{fig:FR+}. This distribution is well approximated by a step function
\begin{gather} 
\frac{d\hspace{.3mm}\text{Num}_{D-q}}{d H_{D-q}^2} \sim \frac{dn}{d H_{D-q}^2} \sim \biggl\{ \begin{array}{ll} 
 \frac{1 }{g H_D^{1+q}} & \textrm{for } H_{D-q} < H_D \\ 
0 & \textrm{for }  H_{D-q} > H_D .
\end{array} 
\label{N1dist}
\end{gather} 
Since the Hubble scale of the parent bounds the Hubble scale of the offspring, $H_{D-q} < H_D$, if you start with a low c.c.~all of the direct descendants have a low c.c.---having low cosmological constant is a heritable trait.

\subsubsection*{$\bm{N=1}$ Freund-Rubin compactifications with $\bm{\Lambda_D \le 0$}: AdS accumulation}

\begin{figure}[htbp] 
   \centering
   \includegraphics[width=\textwidth]{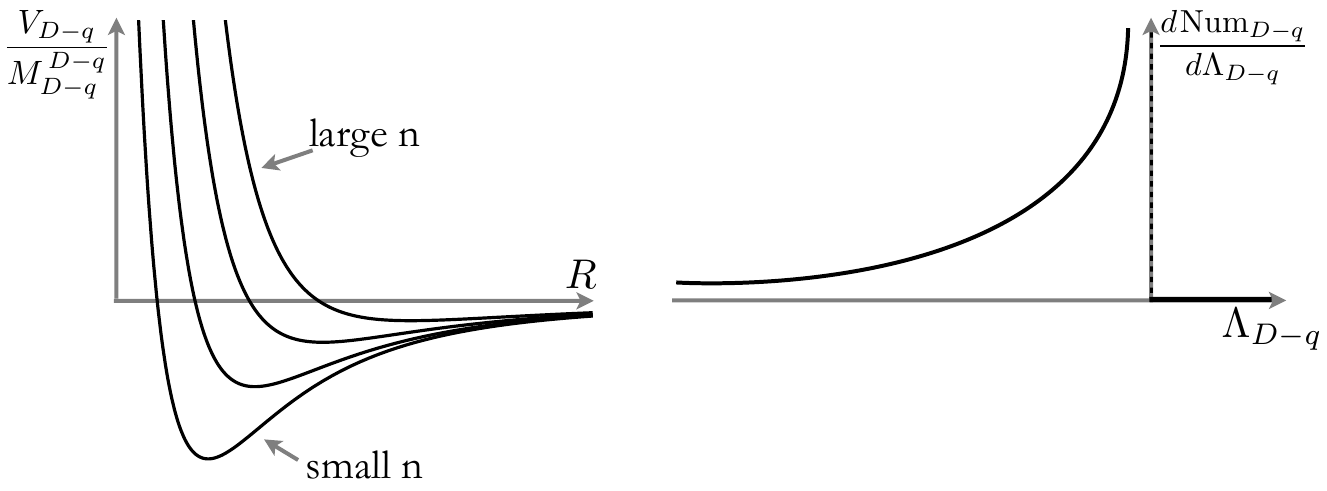} 
   \caption{The $N=1$ FR compactification of $D$-dimensional Minkowski. Left: the effective potential as a function of $R$, plotted for several values of the number of conserved flux units $n$; the minimum is always AdS.  Right: the number density  of AdS$_{D-q}$ vacua as a function of $\Lambda_{D-q}/M_{D-q}^{\,D-q}$.}
   \label{fig:FR-}
\end{figure}

The effective potential and distribution of cosmological constants for $\Lambda_D\le0$ are plotted in Fig.~\ref{fig:FR-}.
The repulsive flux term and the attractive curvature term create a minimum; without positive $\Lambda_D$ to uplift that minimum, it is necessarily AdS$_{D-q}$.
We saw in the last section that the number of $N=1$ minima diverges as $\Lambda_D \rightarrow 0$ from above.   When $\Lambda_D \leq 0$, every value of $n$ gives rise to a minimum and the number of minima is infinite. Unlike the FR model with $\Lambda_D>0$, no matter how many flux units $n$ are wrapped around the extra dimensions, the radion stays stable: the extra dimensions just get larger and larger and the potential gets less and less negative.  Unlike the BP model, no matter how many flux units $n$ are wrapped around the extra dimensions, the energy density stays sub-Planckian: indeed, the extra dimensions grow sufficiently rapidly with $n$ that the flux density $n/R^q$ falls.  There are thus an infinite number of AdS minima in this model, and the cosmological constants of these minima accumulate at zero \cite{Ashok:2003gk}. 

\subsubsection*{$\bm{N = 2}$: de Sitter accumulation}

The BP landscape and $N=1$ compactifications of de Sitter both have smooth distributions of vacua through $\Lambda=0$. We will now see that the same is not true for $N \geq 2$. We begin with the same higher-dimensional action, Eq.~\ref{eq:higherDaction}, except rather than compactifying on a single $q$-sphere, we're going to compactify on $N$ individually-wrapped $q$-spheres.  (A ($1+1$)-dimensional version of this landscape was studied in \cite{Asensio:2012pg,Asensio:2012wt}.  The full spectrum and perturbative stability of these compactifications was studied in \cite{Brown:2013mwa}.)

\newpage

Let's begin with $N=2$.  The ($D-2q)$-dimensional effective potential can be derived by iterating Eq.~\ref{eq:VeffN1}:
\begin{gather}
\frac{V_{D-2q}(R_1,R_2)}{M_{D-2q}^{\,D-2q}}  =  \left( \frac{1}{M_{D-q}^{\,q}
R_2^{\,q}} \right)^{\frac2{D-2q-2}} \left[ \frac{V_{D-q}(R_1)}{M_{D-q}^{\,D-q}}
- \frac{1}{M_{D-q}^{\,2} R_2^{\,2}}  + \frac{g^2 n_2^{\,2}}{M_{{D-q}}^2 R_2^{\,2q}}      \right], \label{eq:effpotproduct}
\end{gather}
where $M_{D-2q}^{\,D-2q-2}=R_2^{\,q}\, M_{D-q}^{D-q-2} = R_2^{\,q}\, R_1^{\,q}\, M_{D}^{D-2}$.  Though not manifest in this form, the potential is symmetric under the exchange of $1\leftrightarrow2$.

\begin{figure}[h!] 
   \centering
   \includegraphics[width=\textwidth]{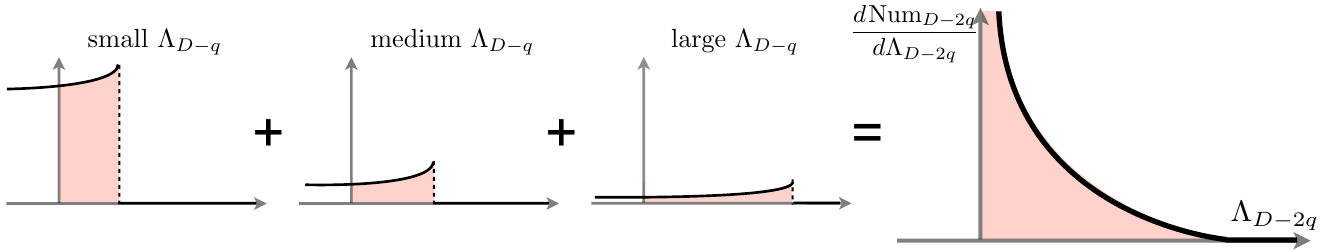} 
   \caption{The number  density of granddaughter vacua with a given value of $\Lambda_{D-2q}$ is the sum of the distributions from each of the $(D-q)$-dimensional daughter vacua. Each daughter vacuum contributes a number of granddaughters inversely proportional to its cosmological constant, with a flat distribution that cuts off at its Hubble scale.  These contributions pile up at $\Lambda_{D-2q}=0$.
   (There's also a divergent number density of AdS vacua for every negative $\Lambda_{D-2q}$.) 
   }
   \label{fig:N2distribution}
\end{figure}

How many dS$_{D-2q}$ minima are there? The easiest way to think about this is in terms of sequential compactification---in terms of first compactifying from $D$ to $D-q$ dimensions, and then compactifying from  $D-q$ to $D-2q$. The first compactification gives rise to a flat range of $\Lambda_{D-q}$'s, as in Fig.~\ref{fig:FR+}. Each of those daughter $\Lambda_{D-q}$'s then gives rise to its own range of granddaughter $\Lambda_{D-2q}$'s with a distribution that is again flat.  The total distribution of granddaughters is given by a convolution, as shown in Fig.~\ref{fig:N2distribution}.  The distribution of Hubbles is
\begin{gather}
\frac{d \hspace{.3mm} \textrm{Num}_{D-2q} }{dH_{D-2q}^{\,2}} 
\sim \int_{H_{D-2q}^{\,2}}^{\infty} \frac{dH_{D-q}^{\,2}}{g H_{D-q}^{q+1}}   \frac{d\,\textrm{Num}_{D-q}}{dH_{D-q}^{\,2}}  
 \sim \int_{H_{D-2q}^{\,2}}^{{H_{D}^{\,2}}} \frac{dH_{D-q}^{\,2}}{g H_{D-q}^{q+1}}  \frac{1}{g H_{D}^{\,q+1}}   \sim \frac1{g^2}\frac1{H_D^{\,q+1}}\frac1{H_{D-2q}^{\,q-1}}\label{eq:Honestep}
\end{gather}
near zero, where, as in Eq.~\ref{N1dist}, we have treated the $N=1$  distribution of Hubbles as a step function.  The $N=2$ distribution is singular at $H_{D-2q}^{\,2}=\Lambda_{D-2q}/M_{D-2q}^{\,D-2q-2}=0$, and the singularity is integrable for $q=2$, logarithmic for $q=3$, and power-law for $q\ge4$. 

\begin{figure}[h!] 
   \centering
   \includegraphics[width=.42\textwidth]{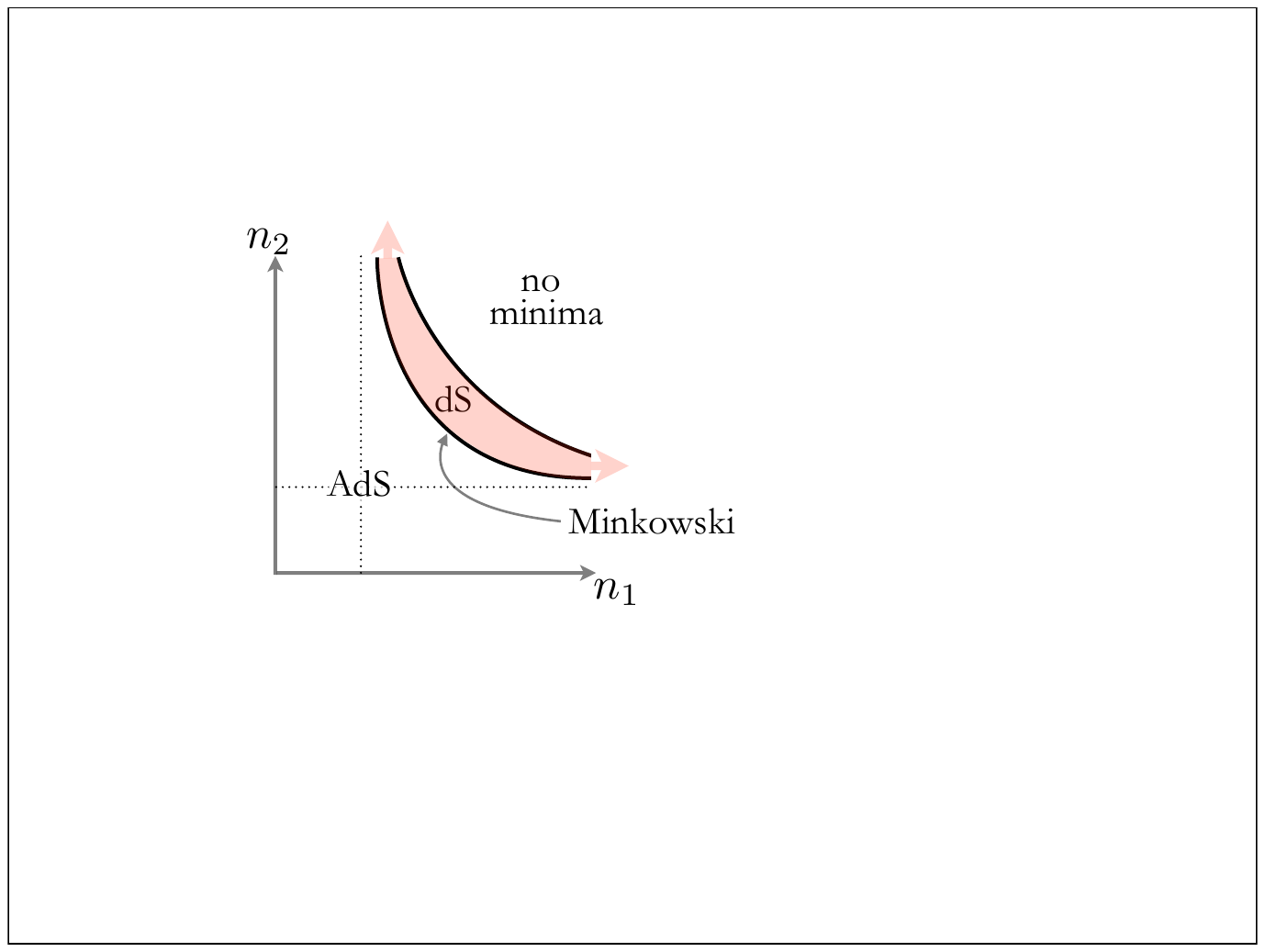} 
   \caption{The phase diagram of $N=2$ compactifications of de Sitter.  The de Sitter minima lie in the infinitely long crescent above the Minkowski line and below the decompactification line.  The area of this strip is a proxy for the number of dS vacua.}
   \label{fig:N2equipotential}
\end{figure}

Another way to think about this singularity is in terms of the phase diagram shown in Fig.~\ref{fig:N2equipotential}. 
While in the BP model the codimension-one surface of Minkowski vacua was a sphere, as in Fig.~\ref{fig:BP}, here the Minkowski surface is a hyperbola.   There is a critical value of $n_1$ indicated by the dotted line to which this hyperbola asymptotes. When $n_1$ is less than this value, the first compactification is to AdS$_{D-q}$ so all values of $n_2$ give rise to minima, but all those minima are AdS$_{D-2q}$.  When $n_1$ is greater than the critical value, the minimum can be either AdS$_{D-2q}$ or dS$_{D-2q}$ or there can be no minimum at all, depending on the value of $n_2$. When $n_1$ is only just above the critical value, $\Lambda_{D-q}$ is only just above zero, and there are de Sitter minima for a large array of $n_2$ all of which have small cosmological constants.  In other words, the large number of dS$_{D-2q}$'s is coming from the `de Sitter tails' on the phase diagram.  The continuous approximation means that we are using area in this phase diagram as a proxy for the number of vacua (the number of grid points).  We will show later that flux quantization generically replaces the infinity with a double-exponentially large number.

\subsubsection*{General $\bm N$: natural selection}

For general $N$ we need to iterate Eq.~\ref{eq:Honestep} $N$ times.  Defining $p\equiv D-Nq$, we have 
\begin{equation}
\frac{d\,\text{Num}_p }{dH_p^{\,2}} \sim \int_{ H^{\,2}_p}  \frac{dH_{p+q}^{\,2}}{g H_{p+q}^{\,1+q} }    \int_{H^2_{p+q}} \frac{dH_{p+2q}^{\,2}}{g H_{p+2q}^{\,1+q} }  \cdots \int^{H^{\,2}_{p + Nq}}_{H^{\,2}_{p + (N-2)q}} \frac{dH_{p+(N-1)q}^{\,2}}{g H_{p+(N-1)q}^{\,1+q} } \frac{1}{g H_{p+Nq}^{\,1+q} }  \sim  \frac1{g^N}\frac1{H_D^{q+1}}\frac1{H_p^{(N-1)(q-1)}} . \label{eq:Hdistribution}  
\end{equation}
The divergence in the distribution of compactified Hubbles is set by $(N-1)(q-1)$.

To find the distribution of cosmological constants $\Lambda_{p}/M_p^{\,p}$, we need to compare these $H_p$'s to the Planck mass $M_p$.  The $p$-dimensional Planck mass is given by 
\begin{equation}
M_p^{p - 2}  \sim R_N^{\,q} R_{N-1}^{\,q} \cdots R_1^{\,q} M_{p + N q}^{p + Nq -2} \sim H_{p+ q}^{-q} H_{p + 2q}^{-q} \cdots H_{p+ Nq}^{-q} M_{p + N q}^{p + Nq -2},
\end{equation}
so that $M_p$ varies from vacuum to vacuum and  depends on the Hubble scale not just of the direct parent, but of all the ancestors; this fact means that the singularity in the distribution of $\Lambda_p/M_p^{\,p}$ will in general be different from the singularity in the distribution of $H_p^{\,2} = \Lambda_p/M_p^{\,p-2}$.  Let's define the ansatz for the distribution of $(\Lambda^{}_p/M_p^{\,p})$'s near zero by 
\begin{equation}
\frac{d\hspace{.3mm}  \textrm{Num}_p}{d(\Lambda_p /M_p^{\,p}) }=\frac{d\hspace{.3mm}  \textrm{Num}_p}{d(H_p^{\,2} /M_p^{\,2}) } \equiv \beta\left(\frac{H_p^{\,2}}{M_p^{\,2}}\right)^{-\alpha}=\beta \left(\frac{\Lambda_p }{M_p^{\,p}} \right)^{-\alpha}.
\end{equation}
We need to modify Eq.~\ref{eq:Hdistribution} by adding in appropriate powers of $M_p$,
\begin{eqnarray}
\frac{d\,\text{Num}_p }{d(H_p^2/M_p^2)} & \sim& M_p^{2 \alpha} \int_{ H^2_p}  \frac{dH_{p+q}^2}{g H_{p+q}^{1+q} }   H_{p + q}^{ {-2q(1-\alpha)/(p-2)}}   \int_{H^2_{p+q}} \frac{dH_{p+2q}^2}{g H_{p+2q}^{1+q} } H_{p +2 q}^{ {-2q(1-\alpha)/(p-2)}}   \ldots  \\
&  & \hspace{1mm}  \ldots  \int_{H^2_{p+(N-2)q}}^{H^2_{p+Nq}} \frac{dH_{p+(N-1)q}^2}{g H_{p+(N-1)q}^{1+q} } H_{p +(N-1) q}^{ {-2q( 1-\alpha)/(p-2)}}  \frac{H_{p + N q}^{ {-2q( 1- \alpha)/(p-2)}}  }{g H_{p+Nq}^{1+q} }  M_{p + Nq}^{2( 1 - \alpha) (p + N q -2) / (p-2)},\nonumber
\end{eqnarray}
which gives a consistency equation for $\alpha = -(N-1)( 1 - \frac{1 +q}{2} - \frac{q (1-\alpha) }{p-2} )$ whose solution is 
\begin{equation}
\boxed{\alpha =  \frac{N-1}{2} \frac{2 + p(q-1)}{D-q-2} } \label{eq:alpharesult} \  \ \ \  \textrm{ with } \ \ \ \beta = \frac{M_D^{2(1-\alpha)(D-2)/(p-2)}}{g^N H_D^{1 + q + 2q(1-\alpha )/(p-2)}} \cdot
\end{equation}
Equation~\ref{eq:alpharesult} gives the distribution of de Sitter vacua near zero for all values of $N\ge1$, $q\ge2$, and $p\ge3$. For $N=1$, the distribution is always flat; but for higher $N$ the distribution diverges at $\Lambda_p=0$: this divergence is integrable for $(N-1)(q-1)\le1$, logarithmic for $(N-1)(q-1)=2$ and power-law for $(N-1)(q-1)\ge3$.  For $p=4$ \& $q=2$, Eq.~\ref{eq:alpharesult} gives
\begin{equation}
\frac{d\,\text{Num}_4}{d(H_4^{\,2}/M_4^{\,2}) }  \sim  \frac{d\,\text{Num}_4}{d(V_4/M_4^{\,4}) }  \sim \left( \frac{ V_4}{M_4^4} \right)^{-\frac{3(N-1)}{2N}},
\end{equation}
consistent with the answer found by a different route in \cite{8dEMmainpaper}.

\subsubsection*{Quantization and infinities}

We have found an infinite volume of $n_i$-space that corresponds to de Sitter minima.  However, the allowed values of $n_i$ are restricted to integer grid points since flux is quantized,  and an infinite volume need not enclose an infinite number of grid points.  Consider the $N=2$ phase diagram in Fig.~\ref{fig:N2equipotential}.  If the coupling $g$ is just right, the grid of allowed $n_i$ will lie exactly along the de Sitter tail, and the infinite flux volume will enclose an infinite number of de Sitter vacua.  However, for generic $g$ the grid will straddle the de Sitter tail and there will only be a finite number of de Sitter vacua. 

While generic $g$ gives a finite number of dS vacua, that finite number is double-exponentially large.  We can see this by again treating the compactification sequentially.
After the first round of compactification, the Hubble of the smallest de Sitter daughter is typically
\begin{equation}
H_{D-q}^{\, 2} \sim \frac{H_D^{\, 2}}{\# \textrm{dS minima}} \sim \frac{H_D^{\, 2}}{H_D^{1-q}/g} \sim g (H_D^{\, 2})^\frac{1+q}{2}.
\end{equation}
The Hubble of the smallest de Sitter daughter of the smallest de Sitter daughter \dots of the smallest de Sitter daughter is therefore typically
\begin{equation}
H_{D - Nq}^2 \sim g^{2  \frac{ (\frac{1 + q}{2})^N - 1}{q-1} } \left( H_D^2 \right)^{\left( \frac{1+q}{2} \right)^N}.
\end{equation}

Since this calculation drastically underestimates the smallness of the smallest positive cosmological constant, this establishes that a generic de Sitter vacuum lies double-exponentially close to $\Lambda_p=0$. 

\subsubsection*{Discussion}

We have shown that if you can construct a $D$-dimensional de Sitter, it will give rise to double-exponentially many lower-dimensional de Sitters, with double-exponentially small cosmological constants. While the KKLT construction of de Sitter \cite{Kachru:2003aw} only works for $D=4$, there are non-critical string theory constructions of de Sitter for any $D$ \cite{Maloney:2002rr}, indicating that our mechanism should operate in the full string-theory landscape. 

Furthermore, while we have illustrated our mechanism with the simplest possible example---$N$ spheres each of dimension $q$---it should operate also in more complicated compactifications. The generalization to spheres of different dimensionality follows trivially by repeated iteration of Eq. 6; the generalization to non-spherical compactifications would be worth investigating further.  
One simple generalization enhances our effect. If instead of a single species of $q$-form flux there are $\mathfrak{n}$ species,  then the Natural Selection effect is even stronger because the reproductive advantage of  low de Sitter minima is even larger: there are $n_{\textrm{max}}^\mathfrak{n}$ ways to wrap flux round a given $q$-sphere such that the total field strength round that sphere stays less than $g^2 n_{\textrm{max}}^2$.

For a landscape in which our mechanism operates, there is no cosmological constant problem, in the sense that typical de Sitter vacua have extremely sub-Planckian cosmological constants. There is however a radion mass problem, in the sense that typical vacua also have extremely large internal dimensions.  Indeed, the lower-dimensional Hubble scale and the KK scale are linked.  In the $N=1$ case, $H_D$ sets the natural scale for both quantities; for general $N$, it is the $(N-1)$st generation's Hubble that sets the natural scale, so typically $m_\text{KK}\sim R_\text{min}^{-1}\sim H_p\sim H_{p+q}$.  The solution to the cosmological constant problem and the introduction of the radion mass problem are two sides of the same coin; it is precisely the large extra dimensions that are diluting away the cosmological constant.  In fact, the extra dimensions dilute away all coupling constants as well, so that the effective four-dimensional theory in one of these vacua is almost completely inert.  This is why quantum corrections do not spoil the accumulation point.

Since the KK scale is so low, a typical de Sitter vacuum in this landscape looks nothing like our own (to say nothing of the absence of the Standard Model).  There are, however, rarer vacua that have a larger KK scale.  How many of these rarer vacua there are depends how we characterize the largeness of the KK scale.  One way to characterize it is to compare $m_\text{KK}$ to $H_p$.  While typical vacua have $H_p\sim H_{p+q}\sim m_\text{KK}$ (they have inherited their parent's Hubble), there are rarer vacua for which $H_p\ll H_{p+q}\sim m_\text{KK}$  so that $m_\text{KK}/H_p$ is large.  Because the distribution of vacua one generation down is flat, as in Eq.~\ref{N1dist}, restricting $m_\text{KK}$ to be greater than some huge number times $H_p$ just multiplies the number density by some overall tiny factor, but leaves the shape of the distribution untouched: the subset of vacua with $M_p\gg m_\text{KK}\gg H_p$ still accumulates at $\Lambda_p=0$.  On the other hand, another way to characterize the largeness of the KK scale, which might be preferable to a four-dimensional effective field theorist, is to compare $m_\text{KK}$ to a four-dimensional energy scale like the TeV scale.  This puts a restriction on the size of the extra dimensions and obliterates the accumulation point.

A related idea for giving rise to a preponderance of low-cosmological-constant vacua was advanced in \cite{Sumitomo:2012wa}. The idea of \cite{Sumitomo:2012wa} is that the product of $N$ uniform independent random variables is peaked at a value that is exponentially small in $N$. The same mathematical fact is also used in this paper---the Weyl factors in the effective potential multiply (see Eq.~\ref{eq:effpotproduct})---and this  was essential to making a small cosmological constant `heritable'. However, the Natural Selection mechanism  makes use of an additional element---that having a small cosmological constant confers to de Sitter vaua a reproductive advantage. It is for this reason that we found the proliferation of low-scale de Sitter vacua to be not exponential but double-exponential.

Non-gravitational physics knows nothing about $\Lambda = 0$ and so cannot give a landscape with a special feature at zero. We have demonstrated a mechanism that picks out $\Lambda=0$ for a fundamentally gravitational reason.
This mechanism can generate a double-exponentially large number of four-dimensional de Sitter vacua and naturally select a small cosmological constant. 

\subsection*{Acknowledgements}
Thanks to Raphael Bousso, Frederik Denef, Shamit Kachru, Liam McAllister, Yasunori Nomura, Eva Silverstein, Paul Steinhardt, Timm Wrase, and Claire Zukowski.


\bibliographystyle{utphys}
\bibliography{mybib.bib}

\end{document}